\documentclass[11pt]{article}

\usepackage[margin=1in]{geometry}
\usepackage{amsmath,amssymb,bm}
\usepackage{booktabs,array,tabularx}
\usepackage{graphicx}
\usepackage{xcolor}
\usepackage{natbib}
\usepackage{etoolbox}
\usepackage[
  colorlinks=true,
  linkcolor=blue,
  citecolor=blue,
  urlcolor=blue
]{hyperref}
\AtBeginEnvironment{tabular*}{\scriptsize\setlength{\tabcolsep}{5pt}}

\newenvironment{keywords}
  {\par\smallskip\noindent\textbf{Keywords:}\ }
  {\par\medskip}

\title{Öpik-type collision frequency for Kozai-driven projectiles: Target bodies on eccentric and inclined orbits}
\author{
\parbox{0.95\textwidth}{\centering
Youpeng Liang,$^{1}$
and Xiaodong Liu$^{1,2}$\thanks{E-mail: liuxd36@mail.sysu.edu.cn}
\\[0.4em]
\small
$^{1}$School of Aeronautics and Astronautics, Shenzhen Campus of Sun Yat-sen University, Shenzhen, Guangdong 518107, China\\
$^{2}$Shenzhen Key Laboratory of Intelligent Microsatellite Constellation, Shenzhen, Guangdong 518107, China
}
}
\date{}

\begin{document}
\label{firstpage}
\maketitle

\begin{abstract}

For high-inclination projectiles undergoing Lidov--Kozai secular evolution, coupled variations in eccentricity and inclination complicate estimates of long-term collision frequencies with a target body. Previous collision-frequency studies for Kozai-driven projectiles have considered target-body orbits that are either circular or confined to the reference plane. The present study considers the general case in which the target body's orbit has both non-zero eccentricity and non-zero inclination. For each complete Lidov--Kozai cycle, the initial relative nodal longitude and the target body's initial argument of periapsis serve as two independent orientation variables. The single-cycle collision frequency is evaluated for each pair of initial values. Under the incommensurability condition considered here, the long-term mean is then obtained by explicitly averaging these values over both orientation variables. For the special cases in which the target body's orbit is circular or lies in the reference plane, the method is shown analytically to reduce to the corresponding formulations of previous studies. Numerical comparisons confirm these reductions. For the general eccentric and inclined case, the resulting collision frequency predicts a projectile survival curve that agrees well with the fraction of projectiles remaining in independent direct dynamical integrations. Finally, it is shown that, for a fixed dynamical setup and under the incommensurability condition, the long-term mean collision frequency associated with a given closed Hamiltonian level curve is independent of both the projectile's initial secular state along that level curve and its initial longitude of ascending node.

\end{abstract}

\section{Introduction}\label{sec:introduction}

For high-inclination projectiles undergoing Lidov--Kozai secular evolution, the eccentricity, inclination, and orbital orientation vary on timescales much longer than the Keplerian orbital period.
These secular variations continually change the geometrical conditions for encounters with a target body.
Consequently, the long-term collision frequency cannot be inferred from a single instantaneous orbital configuration.
Its evaluation must instead combine the secular evolution of the projectile with local collision-probability calculations.

Such local calculations are based on the classical Öpik theory and its subsequent extensions \citep{Opik1951,Wetherill1967,Greenberg1982}.

\citet{vokrouhlicky2012} developed a framework for Kozai-driven projectiles colliding with a target body on a circular orbit in the reference plane.
The secular evolution of the projectile was combined with local collision-probability calculations to obtain the mean collision frequency.

\citet{pokorny2013} extended this treatment to a target body on an eccentric orbit in the reference plane.
The additional dependence on the orientation of the target body's eccentric orbit was incorporated through a weighted integration.

\citet{Liang2026} considered instead a target body on an inclined circular orbit.
In that case, the relative nodal longitude between the projectile's orbit and the target body's orbit provides the additional long-term orientation variable.

These studies treated separately the effects introduced by the eccentricity and inclination of the target body's orbit.
For a target body on an orbit with both non-zero eccentricity and non-zero inclination, however, the two effects must be considered simultaneously.
The orientation of the target body's orbital plane determines the line of intersection between the two orbital planes, whereas its argument of periapsis determines the target body's position and velocity where its orbit crosses this line.
The general problem therefore involves two independent long-term orientation variables.

In the present formulation, these two variables are the relative nodal longitude between the projectile's orbit and the target body's orbit, and the argument of periapsis of the target body's orbit.
For each pair of orientation variables, the collision frequency over one complete Kozai cycle of the projectile is evaluated.
The long-term mean collision frequency is then obtained by averaging over the two orientation variables.
This separates the local collision calculation within a single Kozai cycle from the long-term averaging over orbital orientation.
In the corresponding limiting cases, the formulation reduces to those used in previous studies.

The remainder of this paper is organized as follows.
Section~2 develops the theoretical framework for the long-term mean collision frequency, Section~3 considers its reductions in several special cases, and Section~4 presents the numerical examples and validation.
Section~5 examines whether the projectile's initial secular state along a Hamiltonian level curve affects the long-term mean collision frequency, and Section~6 summarizes the main conclusions.

\section{Theoretical framework for the long-term mean collision frequency}
\label{sec:theoretical-framework}

This section develops the theoretical framework for the long-term mean collision frequency in this general case.
The dynamical variables and time parametrization required for the subsequent collision calculation are introduced first.
The orientation variables that describe the relative orientation of the two orbits at the beginning of a complete Kozai cycle are then defined.
For a given cycle-start orientation, the conditional single-cycle collision frequency over the subsequent complete Kozai cycle is evaluated.
The evolution of these orientation variables between successive Kozai-cycle starts is subsequently determined.
Under the incommensurability condition, this evolution is used to obtain their long-term distribution and the corresponding long-term mean collision frequency.

The single-cycle collision calculation for a given cycle-start orientation builds on the treatment developed by \citet{Liang2026} for a target body on an inclined circular orbit.
When the target body's orbit also has non-zero eccentricity, the orientation of its periapsis becomes an additional independent orientation variable, and the local encounter geometry and Keplerian-motion expressions involving the target body's orbit must be generalized accordingly.
The main text below gives the definitions and main theoretical steps required to construct the long-term mean collision frequency, while the specific local expressions for the general eccentric and inclined target-body orbit are collected in Appendix~\ref{app:general_eccentric_inclined_target}.

\subsection{Dynamical setup of the projectile and target body}
\label{sec:2-1}

\begin{figure*}
\centering
\includegraphics[
    width=0.92\textwidth,
    trim=3pt 3pt 3pt 3pt,
    clip
]{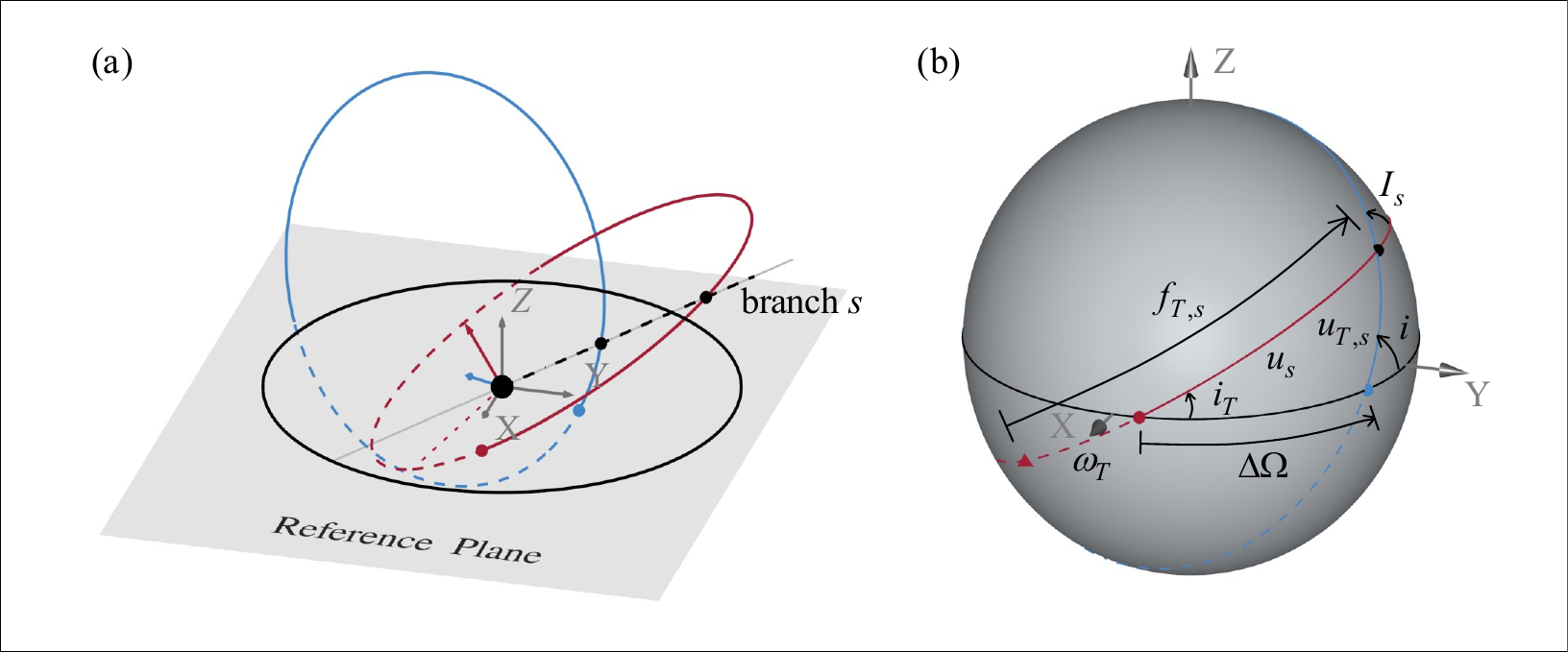}
\caption{
Schematic illustration of the orbital geometry.
Panel~(a) shows the three-dimensional orbital geometry of the system.
The black, red, and blue curves represent the orbits of the perturbing body, the target body, and the projectile, respectively, with the perturbing body's orbital plane defining the reference plane.
The black sphere represents the central body.
The red and blue dots mark the ascending nodes of the target-body orbit and the projectile orbit, respectively, while the red and blue arrows indicate their orbital angular-momentum directions.
The red dotted line indicates the direction from the central body towards the periapsis of the target-body orbit.
The black dashed line indicates branch~\(s\) of the mutual line of nodes between the target-body and projectile orbital planes, and the two black points mark its intersections with the two orbits.
Panel~(b) shows the corresponding orbital-plane geometry on the unit sphere.
The black, red, and blue great circles are the intersections of the perturbing-body, target-body, and projectile orbital planes with the unit sphere, respectively.
The red and blue dots indicate the ascending-node directions of the target-body orbit and the projectile orbit, respectively.
The black dot indicates the direction of branch \(s\), and the red triangle marks the periapsis direction of the target-body orbit.}
\label{fig:dynamical-setup}
\end{figure*}

The dynamical model used here is the quadrupole-order, doubly averaged Lidov--Kozai secular model \citep{Lidov1962,Kozai1962}.
The system consists of a central body, a perturbing body, a projectile, and a target body.
The perturbing body moves on a circular Keplerian orbit about the central body, and its orbital plane is adopted as the reference plane, relative to which all inclinations used below are measured.
The projectile is treated as a massless test particle, and its physical size is neglected.
Its long-term orbital evolution is driven by the quadrupole-order secular perturbation induced by the perturbing body.
The target body's orbit is described by prescribed Keplerian orbital elements, with its orbital orientation evolving in time as specified below.
The dynamical system and its orbital geometry are illustrated schematically in Figure~\ref{fig:dynamical-setup}.

The derivation of the above secular model is not repeated here; only the dynamical variables, conserved quantities, secular evolution equations, and time parametrization required for the subsequent collision-frequency calculation are given.
Let \(a,e,i,\omega,\Omega\) denote the semimajor axis, eccentricity, inclination, argument of periapsis, and longitude of ascending node of the projectile's orbit, respectively.
Within the quadrupole-order secular approximation, the semimajor axis \(a\) is conserved, as is the Kozai constant
\begin{equation}
c=\sqrt{1-e^2}\cos i .
\label{eq:kozai-constant}
\end{equation}

To describe the secular variations of the eccentricity and periapsis orientation, the nonsingular eccentricity variables are introduced as
\begin{equation}
k=e\cos\omega,\qquad
h=e\sin\omega,
\label{eq:kh-definition}
\end{equation}
and
\begin{equation}
g\equiv\sqrt{1-e^2}
=\sqrt{1-k^2-h^2}.
\label{eq:g-definition}
\end{equation}
For \(e>0\), the corresponding orbital elements can be recovered from
\begin{equation}
e=\sqrt{k^2+h^2},\qquad
\omega=\operatorname{atan2}(h,k),\qquad
\cos i=\frac{c}{g}.
\end{equation}

Adopting the same normalization as \citet{vokrouhlicky2012} and \citet{Liang2026}, the quadrupole-order secular Hamiltonian of the projectile is written as
\begin{equation}
\begin{aligned}
H(k,h;c)
={}
\left[2+3(k^2+h^2)\right]
\left(\frac{3c^2}{g^2}-1\right)
+
15(k^2-h^2)
\left(1-\frac{c^2}{g^2}\right).
\end{aligned}
\label{eq:kozai-hamiltonian}
\end{equation}

Using the dimensionless secular time \(\tau\), the secular evolution of \(k\) and \(h\) is governed by
\begin{equation}
\begin{aligned}
\frac{{\rm d}k}{{\rm d}\tau}
&=
-\frac{\partial H}{\partial h}
=
\frac{12h}{g^4}
\left[
3g^4-5c^2(1-k^2)
\right],
\\
\frac{{\rm d}h}{{\rm d}\tau}
&=
\frac{\partial H}{\partial k}
=
\frac{12k}{g^4}
\left(
2g^4+5c^2h^2
\right).
\end{aligned}
\label{eq:secular-evolution-kh}
\end{equation}

Because \(c\) is conserved and the Hamiltonian has no explicit dependence on \(\tau\), \(H\) remains constant along the secular evolution.
For a given initial state \((k_0,h_0)\), let
\begin{equation}
H_0=H(k_0,h_0;c).
\end{equation}
The secular evolution of the projectile in the \((k,h)\) plane is then confined to the Hamiltonian level curve
\begin{equation}
H(k,h;c)=H_0.
\label{eq:hamiltonian-level-curve}
\end{equation}
For the Kozai-resonant orbits considered here, this level curve is closed, and the projectile evolves periodically along it.
The Hamiltonian level curve specifies the secular states in the \((k,h)\) plane that are encountered during a complete Kozai cycle, while the secular evolution equations and the associated time parametrization determine the order in which these states are traversed and their corresponding physical times.

With the same normalized Hamiltonian, the secular evolution of the projectile's longitude of ascending node is
\begin{equation}
\frac{{\rm d}\Omega}{{\rm d}\tau}
=
-\frac{1}{g}
\frac{\partial H}{\partial c}
=
-\frac{12c}{g^3}
\left(
1-k^2+4h^2
\right).
\label{eq:dOmega-dtau}
\end{equation}
Thus, once the initial secular state is specified, equations~\eqref{eq:secular-evolution-kh} and~\eqref{eq:dOmega-dtau} determine the secular evolution of \(k(\tau)\), \(h(\tau)\), and \(\Omega(\tau)\).

The dimensionless secular time \(\tau\) is related to the physical time \(t\) through
\begin{equation}
{\rm d}t
=
\frac{16}{\gamma_\star g}\,
{\rm d}\tau,
\label{eq:time-reparametrization}
\end{equation}
where \(\gamma_\star\) is the characteristic frequency scale associated with the quadrupole-order secular perturbation.
For the circular orbit of the perturbing body adopted here,
\begin{equation}
\gamma_\star
=
\frac{\mu_P}{a_P^3}
\sqrt{\frac{a^3}{\mu_0}},
\label{eq:gamma-star}
\end{equation}
where \(\mu_0\) and \(\mu_P\) are the gravitational parameters of the central body and the perturbing body, respectively, and \(a_P\) is the semimajor axis of the perturbing body's orbit.

Let \(\tau_0\) denote the reference origin of the dimensionless secular time, and let the corresponding physical time be \(t_0=t(\tau_0)\).
The physical time associated with an arbitrary secular time \(\tau\) is then
\begin{equation}
t(\tau)
=
t_0+
\int_{\tau_0}^{\tau}
\frac{16}
{\gamma_\star g(\tau')}
\,{\rm d}\tau'.
\label{eq:time-mapping}
\end{equation}
Thus, each value of the dimensionless secular time \(\tau\) is associated with a unique physical time \(t(\tau)\).

Let \(\tau_{\rm K}\) denote the dimensionless time required for the projectile to complete one full Kozai cycle along a closed Hamiltonian level curve.
The corresponding physical duration is
\begin{equation}
T_{\rm K}
=
t(\tau_0+\tau_{\rm K})-t(\tau_0)
=
\int_{\tau_0}^{\tau_0+\tau_{\rm K}}
\frac{16}
{\gamma_\star g(\tau)}
\,{\rm d}\tau.
\label{eq:Kozai-period}
\end{equation}

Let the orbital elements of the target body be
\begin{equation}
a_T,\quad e_T,\quad i_T,\quad \Omega_T,\quad \omega_T,
\end{equation}
where \(a_T\), \(e_T\), and \(i_T\) are the semimajor axis, eccentricity, and inclination of the target body's orbit relative to the reference plane, respectively, while \(\Omega_T\) and \(\omega_T\) are its longitude of ascending node and argument of periapsis.
The mutual gravitational interaction between the target body and the projectile is neglected, and the target body's orbit is prescribed through its orbital elements and their specified orientation evolution.

Over the secular timescales considered here, \(a_T\), \(e_T\), and \(i_T\) are held fixed, while the longitude of ascending node and the argument of periapsis of the target body's orbit are prescribed to vary with time.
Using the reference time \(t_0=t(\tau_0)\) defined above, let
\begin{equation}
\Omega_{T,0}=\Omega_T(t_0),\qquad
\omega_{T,0}=\omega_T(t_0).
\end{equation}
Their time evolution is then written as
\begin{equation}
\Omega_T(t)
=
\Omega_{T,0}
+
\dot{\Omega}_T
\left(t-t_0\right),
\label{eq:target-node-evolution}
\end{equation}
and
\begin{equation}
\omega_T(t)
=
\omega_{T,0}
+
\dot{\omega}_T
\left(t-t_0\right),
\label{eq:target-periapsis-evolution}
\end{equation}
where
\begin{equation}
\dot{\Omega}_T
=
\frac{{\rm d}\Omega_T}{{\rm d}t},
\qquad
\dot{\omega}_T
=
\frac{{\rm d}\omega_T}{{\rm d}t},
\end{equation}
are the signed rates of change of the target body's longitude of ascending node and argument of periapsis, respectively.
Both rates are taken to be constant over the timescales considered here and may be positive, negative, or zero.

Along the dimensionless secular-time parametrization of the projectile, the following shorthand is used:
\begin{equation}
\Omega_T(\tau)\equiv\Omega_T[t(\tau)],
\qquad
\omega_T(\tau)\equiv\omega_T[t(\tau)].
\label{eq:target-orientation-tau}
\end{equation}
This allows the relative orbital orientation of the projectile and the target body to be specified at the same secular time \(\tau\).

Finally, let \(R_T\) denote the geometrical radius of the target body.
Gravitational focusing is not included in the collision cross-section, and the collision radius used below is therefore taken to be \(R=R_T\).

\subsection{Orientation variables at the start of a Kozai cycle}
\label{sec:2-2}

Within the dynamical setup defined above, the instantaneous orbital geometry between the projectile and the target body also depends on the relative orientation of their orbits.
For the general case considered here, with \(e_T\neq0\) and \(i_T\neq0\), two angular variables are required to describe this relative orientation.

The relative nodal longitude between the projectile's orbit and the target body's orbit is defined as
\begin{equation}
\Delta\Omega(\tau)
=
\Omega(\tau)-\Omega_T(\tau).
\label{eq:relative-node}
\end{equation}
Here, \(\Omega(\tau)\) evolves according to equation~\eqref{eq:dOmega-dtau}.
The corresponding \(\Omega_T(\tau)\) is obtained from equation~\eqref{eq:target-node-evolution} using the time mapping in equation~\eqref{eq:time-mapping} and the shorthand defined in equation~\eqref{eq:target-orientation-tau}.
Thus, \(\Delta\Omega(\tau)\) describes the relative nodal orientation of the two orbital planes.

Because \(e_T\neq0\), the target body's orbit is not rotationally symmetric within its own orbital plane.
For a given spatial direction, both the radial distance of the target body's orbit from the central body and the corresponding local orbital geometry depend on the orientation of that direction relative to the target body's periapsis.
Consequently, \(\Delta\Omega(\tau)\) alone is insufficient to specify the relative orientation of the two orbits, and the target body's argument of periapsis \(\omega_T(\tau)\) must also be retained.
The corresponding \(\omega_T(\tau)\) is obtained analogously from equation~\eqref{eq:target-periapsis-evolution} using the time mapping in equation~\eqref{eq:time-mapping} and the shorthand in equation~\eqref{eq:target-orientation-tau}.
Both \(\Delta\Omega\) and \(\omega_T\) are henceforth understood as angular variables modulo \(2\pi\).

The collision-frequency calculation is organized in complete Kozai cycles.
Starting from the reference secular time \(\tau_0\), the beginning and end of the \(n\)th complete Kozai cycle are defined as
\begin{equation}
\tau_{n-1}
=
\tau_0+(n-1)\tau_{\rm K},
\qquad
\tau_n
=
\tau_0+n\tau_{\rm K},
\qquad
n=1,2,3,\ldots ,
\label{eq:cycle-boundaries}
\end{equation}
where \(\tau_{\rm K}\) is the dimensionless duration of one complete Kozai cycle defined in equation~\eqref{eq:Kozai-period}.

Because \(\tau_{\rm K}\) corresponds to one complete traversal of the same closed Hamiltonian level curve, the projectile returns to the same \((k,h)\) state at the beginning of every Kozai cycle,
\begin{equation}
k(\tau_{n-1})=k(\tau_0),
\qquad
h(\tau_{n-1})=h(\tau_0),
\qquad
n=1,2,3,\ldots .
\label{eq:cycle-start-kh-return}
\end{equation}
Thus, once the Hamiltonian level curve and the initial state \((k_0,h_0)\) on that level curve are fixed, successive Kozai cycles begin from the same \((k,h)\) state and differ only through their cycle-start orbital orientations.

The two orientation variables at the beginning of the \(n\)th Kozai cycle are collected into
\begin{equation}
\mathbf{x}_{n-1}
=
\left(
\Delta\Omega_{n-1},
\omega_{T,n-1}
\right),
\label{eq:cycle-start-orientation}
\end{equation}
where
\begin{equation}
\Delta\Omega_{n-1}
=
\Delta\Omega(\tau_{n-1}),
\qquad
\omega_{T,n-1}
=
\omega_T(\tau_{n-1}).
\end{equation}

It is important that \(\mathbf{x}_{n-1}\) specifies only the relative orbital orientation at the beginning of the \(n\)th Kozai cycle.
It does not imply that \(\Delta\Omega\) or \(\omega_T\) remains fixed during the subsequent cycle.
Over \(\tau\in[\tau_{n-1},\tau_n]\), \(k(\tau)\) and \(h(\tau)\) evolve according to equation~\eqref{eq:secular-evolution-kh}, while \(\Omega(\tau)\) evolves according to equation~\eqref{eq:dOmega-dtau}.
The target-body quantities \(\Omega_T(\tau)\) and \(\omega_T(\tau)\) are obtained from equations~\eqref{eq:target-node-evolution} and~\eqref{eq:target-periapsis-evolution}, respectively, through the time mapping in equation~\eqref{eq:time-mapping}.
Accordingly, once the cycle-start orientation \(\mathbf{x}_{n-1}\) is specified, the dynamical evolution defined in Section~\ref{sec:2-1} determines the relative orbital geometry throughout the subsequent complete Kozai cycle.

\subsection{Conditional single-cycle collision frequency for a given cycle-start orientation}
\label{sec:2-3}

For a fixed Hamiltonian level curve and a fixed initial state \((k_0,h_0)\) on that level curve, specifying the cycle-start orientation \(\mathbf{x}_{n-1}\) determines, through the dynamical evolution defined above, the relative orbital geometry over the interval \(\tau\in[\tau_{n-1},\tau_n]\).
The corresponding conditional collision frequency over that complete Kozai cycle is then evaluated as follows.

At any secular time \(\tau\), the intersection of the projectile's orbital plane with the target body's orbital plane defines the mutual line of nodes.
Its two opposite directions form the two mutual-node branches, labelled by \(s=\pm1\).
Let \(r_s(\tau)\) and \(r_{T,s}(\tau)\) denote the radial distances from the central body to the projectile's orbit and the target body's orbit, respectively, along mutual-node branch \(s\).
Their signed radial-distance difference is defined as
\begin{equation}
G_s(\tau)
=
r_s(\tau)-r_{T,s}(\tau),
\qquad
s=\pm1 .
\label{eq:signed-radial-difference}
\end{equation}
The explicit expressions for \(r_s\), \(r_{T,s}\), and the associated local encounter geometry for a general eccentric and inclined target-body orbit are given in Appendix~\ref{app:general_eccentric_inclined_target}.

An exact orbital-intersection configuration along mutual-node branch \(s\) occurs when
\begin{equation}
G_s(\tau_{s,w})=0.
\label{eq:exact-intersection-root}
\end{equation}
Here, \(\tau_{s,w}\) denotes the \(w\)th exact orbital-intersection root on branch \(s\) within the \(n\)th Kozai cycle.
A given branch may contain no such root, one root, or multiple roots during a complete Kozai cycle.

Because the target body has a finite collision radius \(R\), each exact orbital-intersection root \(\tau_{s,w}\) is associated with a local time window around the root.
Within this window, the secular evolution moves the system away from the exact intersection configuration, while the local minimum distance between the two orbits remains no greater than the collision radius \(R\).
Let \(\Delta t_{s,w}\) denote the duration of this window in physical time.
The corresponding slow-variable probability factor is defined as
\begin{equation}
P_{1,s,w}
=
\frac{\Delta t_{s,w}}{T_{\rm K}} .
\label{eq:P1-definition}
\end{equation}
Thus, \(P_{1,s,w}\) represents the fraction of one complete Kozai cycle for which the secular evolution keeps the system within the local collision geometry associated with that root.
The construction of the local time window for the general eccentric and inclined target-body orbit is given in Appendix~\ref{app:general_eccentric_inclined_target}.

For a given local slow-variable state, the fast Keplerian motion of the target body must also be taken into account.
For the root \(\tau_{s,w}\), let \(P_{2,s,w}\) denote the corresponding fast-phase probability factor.
This factor gives the probability that, once the local orbital-approach condition around the root is satisfied, the target body reaches the local encounter region with a suitable fast Keplerian phase for an actual collision to occur.
Its evaluation is based on the local orbital geometry and velocity relations at the root.
The finite collision radius determines the range of target-body fast phases that can lead to collision, and the target body's Keplerian motion is then used to convert this allowed range into the corresponding probability.
The required local geometry, velocity relations, and explicit expressions for the general eccentric and inclined target-body orbit are given in Appendix~\ref{app:general_eccentric_inclined_target}.

The dimensionless root-level collision contribution associated with the \(w\)th exact orbital-intersection root on mutual-node branch \(s\) is therefore defined as
\begin{equation}
\Pi_{s,w}
=
P_{1,s,w}P_{2,s,w}.
\label{eq:root-contribution}
\end{equation}

Let the Keplerian orbital period of the projectile about the central body be
\begin{equation}
T_{\rm orb}
=
2\pi
\sqrt{\frac{a^3}{\mu_0}} .
\label{eq:projectile-orbital-period}
\end{equation}
Since \(a\) is conserved in the secular model adopted here, \(T_{\rm orb}\) is constant.

Summing the contributions from all exact orbital-intersection roots on both mutual-node branches during the \(n\)th complete Kozai cycle gives the conditional single-cycle collision frequency for the cycle-start orientation \(\mathbf{x}_{n-1}\),
\begin{equation}
K(\mathbf{x}_{n-1})
=
\frac{1}{T_{\rm orb}}
\sum_{s=\pm1}
\sum_w
\Pi_{s,w}
=
\frac{1}{T_{\rm orb}}
\sum_{s=\pm1}
\sum_w
P_{1,s,w}P_{2,s,w}.
\label{eq:single-cycle-frequency}
\end{equation}
Accordingly, \(K(\mathbf{x}_{n-1})\) has dimensions of inverse time. It represents the conditional collision frequency obtained from all local orbital-intersection configurations encountered during the complete Kozai cycle that begins with the cycle-start orientation \(\mathbf{x}_{n-1}\).

The slow-variable--fast-phase decomposition used above follows the basic construction of \citet{Liang2026}.
The local orbital geometry and Keplerian-motion expressions required when the target body's orbit has both non-zero eccentricity and non-zero inclination are collected in Appendix~\ref{app:general_eccentric_inclined_target}.

\subsection{Evolution of the orientation variables between successive Kozai-cycle starts}
\label{sec:2-4}

To obtain the long-term mean collision frequency from the conditional single-cycle frequency \(K(\mathbf{x}_{n-1})\), the evolution of the orientation variables between successive Kozai-cycle starts must be determined.

According to equation~\eqref{eq:cycle-boundaries}, the \(n\)th Kozai cycle spans the interval \([\tau_{n-1},\tau_n]\).
Because \(k(\tau)\) and \(h(\tau)\) evolve periodically along the same closed Hamiltonian level curve, they return to the same secular state after one complete Kozai cycle.
The projectile's longitude of ascending node \(\Omega\), however, generally accumulates a non-zero net change.
This nodal increment over one Kozai cycle is defined as
\begin{equation}
\delta_{\Omega}
\equiv
\Omega(\tau_n)-\Omega(\tau_{n-1})
=
\int_{\tau_{n-1}}^{\tau_n}
\frac{{\rm d}\Omega}{{\rm d}\tau}\,
{\rm d}\tau ,
\label{eq:projectile-node-increment}
\end{equation}
where \({\rm d}\Omega/{\rm d}\tau\) is given by equation~\eqref{eq:dOmega-dtau}.
Because the right-hand side of equation~\eqref{eq:dOmega-dtau} depends only on the periodic variables \(k(\tau)\), \(h(\tau)\), and the conserved quantity \(c\), the same closed Hamiltonian level curve gives the same value of \(\delta_{\Omega}\) for every complete Kozai cycle.

Over the corresponding physical duration \(T_{\rm K}\), the target body's longitude of ascending node and argument of periapsis change by
\begin{equation}
\dot{\Omega}_T T_{\rm K},
\qquad
\dot{\omega}_T T_{\rm K},
\end{equation}
respectively, according to equations~\eqref{eq:target-node-evolution} and~\eqref{eq:target-periapsis-evolution}.
The increments of the relative nodal longitude and the target body's argument of periapsis between two successive Kozai-cycle starts are therefore
\begin{equation}
\delta_{\Delta\Omega}
=
\delta_{\Omega}
-
\dot{\Omega}_T T_{\rm K},
\qquad
\delta_{\omega_T}
=
\dot{\omega}_T T_{\rm K}.
\label{eq:orientation-increments}
\end{equation}

The orientation variables at the beginning of the next Kozai cycle consequently satisfy
\begin{equation}
\mathbf{x}_{n}
=
\mathbf{x}_{n-1}
+
\left(
\delta_{\Delta\Omega},
\delta_{\omega_T}
\right)
\pmod{2\pi}.
\label{eq:orientation-map}
\end{equation}
Thus, the sequence of cycle-start orientations is generated by a fixed translation in the two-dimensional orientation space.

The incommensurate case considered here is that in which the two angular increments and \(2\pi\) satisfy no non-trivial integer relation.
Explicitly, for any
\begin{equation}
(z_{\Delta\Omega},z_{\omega_T})
\in
\mathbb{Z}^2\setminus\{(0,0)\}
\end{equation}
and any \(z_0\in\mathbb{Z}\),
\begin{equation}
z_{\Delta\Omega}\,
\delta_{\Delta\Omega}
+
z_{\omega_T}\,
\delta_{\omega_T}
\neq
2\pi z_0 .
\label{eq:incommensurability}
\end{equation}
Under this condition, the sequence generated by equation~\eqref{eq:orientation-map} is equidistributed over the two-dimensional orientation space in the long-term limit.
This property allows the long-term discrete average over successive Kozai cycles to be replaced by a continuous average over the orientation space, as shown in the next subsection.

\subsection{Long-term orientation distribution and mean collision frequency}
\label{sec:2-5}

Equation~\eqref{eq:single-cycle-frequency} gives the conditional single-cycle collision frequency \(K(\mathbf{x}_{n-1})\) associated with the complete Kozai cycle that begins with orientation \(\mathbf{x}_{n-1}\).
Because every complete Kozai cycle on the same closed Hamiltonian level curve has the same physical duration \(T_{\rm K}\), the long-term time average over successive cycles is equivalent to the arithmetic mean of their conditional single-cycle collision frequencies.
The mean collision frequency over \(N\) successive complete Kozai cycles is therefore
\begin{equation}
\Gamma^{(N)}
=
\frac{1}{N}
\sum_{n=1}^{N}
K(\mathbf{x}_{n-1}) .
\label{eq:finite-cycle-average}
\end{equation}
The long-term mean collision frequency is defined as
\begin{equation}
\Gamma
=
\lim_{N\rightarrow\infty}
\Gamma^{(N)} .
\label{eq:long-term-frequency-limit}
\end{equation}

To express equation~\eqref{eq:long-term-frequency-limit} as a continuous average over the orientation space, consider an arbitrary point in the two-dimensional orientation space,
\begin{equation}
\mathbf{x}'
=
\left(
\Delta\Omega',
\omega_T'
\right),
\qquad
0\leq\Delta\Omega'<2\pi,
\qquad
0\leq\omega_T'<2\pi .
\label{eq:generic-orientation}
\end{equation}
The corresponding differential element is defined as
\begin{equation}
{\rm d}\mathbf{x}'
\equiv
{\rm d}\Delta\Omega'\,
{\rm d}\omega_T'.
\end{equation}
Because both orientation variables are defined modulo \(2\pi\), the two-dimensional orientation space may be represented by the fundamental domain
\begin{equation}
\mathcal{X}
=
[0,2\pi)\times[0,2\pi),
\label{eq:orientation-space}
\end{equation}
with opposite boundaries identified periodically.

With the Hamiltonian level curve and the chosen initial state \((k_0,h_0)\) on it fixed, \(K(\mathbf{x}')\) is evaluated as a function of the cycle-start orientation variables \(\mathbf{x}'=(\Delta\Omega',\omega_T')\), while all other parameters retain the values specified above.
Only \(\mathbf{x}'\) is varied over \(\mathcal{X}\), following the construction described in Section~\ref{sec:2-3}.
Thus, \(K\) may be regarded as a function defined over the two-dimensional orientation space \(\mathcal{X}\).

Under the incommensurability condition in equation~\eqref{eq:incommensurability}, the sequence of successive cycle-start orientations generated by equation~\eqref{eq:orientation-map} is equidistributed over \(\mathcal{X}\) in the long-term limit.
Its long-term probability density over the orientation space is therefore
\begin{equation}
W(\Delta\Omega',\omega_T')
=
\frac{1}{(2\pi)^2}.
\label{eq:orientation-density}
\end{equation}
Accordingly, the long-term discrete average in equation~\eqref{eq:long-term-frequency-limit} can be written equivalently as the continuous average over the orientation space,
\begin{equation}
\Gamma
=
\int_{\mathcal{X}}
K(\mathbf{x}')
W(\mathbf{x}')
\,{\rm d}\mathbf{x}'
=
\frac{1}{(2\pi)^2}
\int_0^{2\pi}
\int_0^{2\pi}
K(\Delta\Omega',\omega_T')
\,{\rm d}\Delta\Omega'
\,{\rm d}\omega_T' .
\label{eq:long-term-orientation-average}
\end{equation}

Equation~\eqref{eq:long-term-orientation-average} is the fundamental expression for the long-term mean collision frequency in the general case of an eccentric and inclined target-body orbit.
It separates the collision calculation over one complete Kozai cycle from the long-term distribution of cycle-start orientations: for each prescribed pair of orientation variables, the corresponding conditional single-cycle collision frequency is first evaluated, and the resulting values are then averaged over all possible cycle-start orientations.

The intrinsic collision probability is further defined as
\begin{equation}
p
=
\frac{\Gamma}{R^2}.
\label{eq:intrinsic-collision-probability}
\end{equation}
Thus, \(p\) has dimensions of inverse area per unit time.

\section{Theoretical reductions for special target-body orbits}
\label{sec:limits}

The previous section established the framework for the long-term mean collision frequency when the target body's orbit has both non-zero eccentricity and non-zero inclination.
In the general case, the conditional single-cycle collision frequency \(K(\mathbf{x}')\) depends on the two cycle-start orientation variables
\(
\mathbf{x}'
=
\left(
\Delta\Omega',
\omega_T'
\right),
\)
and the long-term mean collision frequency is obtained by averaging \(K(\mathbf{x}')\) over the two-dimensional orientation space \(\mathcal{X}\).
For certain special target-body orbital geometries, the additional orbital symmetry causes one or both orientation variables to lose independent physical significance.
The conditional single-cycle collision frequency then depends on fewer independent orientation variables, and the corresponding long-term averaging simplifies accordingly.
Three such cases are considered below, showing how the general formulation reduces to the corresponding collision-frequency formulations used in previous studies.

\subsection{\texorpdfstring{Special case with \(e_T=0\) and \(i_T=0\)}{Special case with eT = 0 and iT = 0}}
\label{sec:limit-et0-it0}

In this case, the target body moves on a circular orbit in the reference plane.
Because \(e_T=0\), the target body's orbit has no physically distinct periapsis direction, while \(i_T=0\) means that its ascending-node direction likewise has no independent physical significance.
Moreover, the circular target-body orbit is rotationally symmetric within the reference plane.
Consequently, neither the cycle-start relative nodal longitude \(\Delta\Omega'\) nor the target body's argument of periapsis \(\omega_T'\) affects the local orbital-intersection geometry during the subsequent complete Kozai cycle.
The corresponding exact orbital-intersection configurations and their collision contributions are therefore independent of the cycle-start orientation \(\mathbf{x}'\), and the conditional single-cycle collision frequency can be written as
\begin{equation}
K(\Delta\Omega',\omega_T')
=
K.
\end{equation}

Substituting this relation into equation~\eqref{eq:long-term-orientation-average} gives
\begin{equation}
\Gamma
=
\frac{1}{(2\pi)^2}
\int_0^{2\pi}
\int_0^{2\pi}
K
\,{\rm d}\Delta\Omega'
\,{\rm d}\omega_T'
=
K.
\end{equation}
Thus, in the limit \(e_T=0\) and \(i_T=0\), the conditional single-cycle collision frequency is independent of both cycle-start orientation variables, and the average over the two-dimensional orientation space reduces to \(\Gamma=K\).
This result is consistent with the corresponding collision-frequency formulation adopted by \citet{vokrouhlicky2012} for a circular target-body orbit in the reference plane.

\subsection{\texorpdfstring{Special case with \(e_T=0\) and \(i_T\neq0\)}{Special case with eT = 0 and iT != 0}}
\label{sec:limit-et0-itneq0}

In this case, the target body moves on a circular orbit inclined relative to the reference plane.
Because \(i_T\neq0\), the orientation of the target body's orbital plane remains geometrically meaningful.
The cycle-start relative nodal longitude \(\Delta\Omega_{n-1}\) between the projectile's orbit and the target body's orbit therefore continues to affect the local orbital-intersection geometry during the subsequent complete Kozai cycle.
On the other hand, because \(e_T=0\), the target body's orbit has no physically distinct periapsis direction, and its argument of periapsis \(\omega_{T,n-1}\) no longer affects the collision geometry.
The conditional single-cycle collision frequency therefore reduces from the general form
\begin{equation}
K(\mathbf{x}_{n-1})
=
K(\Delta\Omega_{n-1},\omega_{T,n-1})
\end{equation}
to
\begin{equation}
K(\Delta\Omega_{n-1}).
\end{equation}

For the integration variables \((\Delta\Omega',\omega_T')\), the corresponding relation is
\begin{equation}
K(\Delta\Omega',\omega_T')
=
K(\Delta\Omega').
\end{equation}
Substituting this relation into equation~\eqref{eq:long-term-orientation-average} gives
\begin{equation}
\Gamma
=
\frac{1}{(2\pi)^2}
\int_0^{2\pi}
\int_0^{2\pi}
K(\Delta\Omega')
\,{\rm d}\Delta\Omega'
\,{\rm d}\omega_T'
=
\frac{1}{2\pi}
\int_0^{2\pi}
K(\Delta\Omega')
\,{\rm d}\Delta\Omega'.
\end{equation}
Thus, in the limit \(e_T=0\) and \(i_T\neq0\), the general two-dimensional orientation space reduces to a one-dimensional orientation space described solely by the relative nodal longitude \(\Delta\Omega'\).
The long-term mean collision frequency correspondingly reduces to an average over this one-dimensional orientation space.

The relative nodal longitudes at successive Kozai-cycle starts satisfy
\begin{equation}
\Delta\Omega_n
=
\Delta\Omega_{n-1}
+
\delta_{\Delta\Omega}
\pmod{2\pi}.
\end{equation}
For this reduced one-dimensional problem, replacing the long-term cycle average by a uniform average over \(\Delta\Omega'\) requires only that \(\delta_{\Delta\Omega}\) be incommensurate with \(2\pi\).
Equivalently, there are no integers \(m\neq0\) and \(q\) such that
\begin{equation}
m\,\delta_{\Delta\Omega}
=
2\pi q.
\end{equation}
Under this condition, the successive cycle-start values \(\Delta\Omega_{n-1}\) are equidistributed over the corresponding one-dimensional orientation space in the long-term limit.

This continuous average over \(\Delta\Omega'\) can also be related directly to the multi-cycle discrete average used by \citet{Liang2026} for an inclined circular target-body orbit.
Using the cycle indexing defined above, the cumulative mean collision frequency over the first \(N\) consecutive Kozai cycles is
\begin{equation}
\Gamma^{(N)}
=
\frac{1}{N}
\sum_{n=1}^{N}
K(\Delta\Omega_{n-1}).
\end{equation}
As \(N\rightarrow\infty\), the equidistribution of \(\Delta\Omega_{n-1}\) over the corresponding one-dimensional orientation space gives
\begin{equation}
\lim_{N\rightarrow\infty}
\Gamma^{(N)}
=
\frac{1}{2\pi}
\int_0^{2\pi}
K(\Delta\Omega')
\,{\rm d}\Delta\Omega'
=
\Gamma.
\end{equation}
Thus, the discrete average over consecutive Kozai cycles used by \citet{Liang2026} and the continuous average over \(\Delta\Omega'\) used here are two equivalent representations of the same long-term mean collision frequency.

\subsection{\texorpdfstring{Special case with \(e_T\neq0\) and \(i_T=0\)}{Special case with eT != 0 and iT = 0}}
\label{sec:limit-etneq0-it0}

In this case, the target body moves on an eccentric orbit in the reference plane.
Because \(i_T=0\), the target body's orbital plane coincides with the reference plane, and its ascending-node direction no longer has independent physical significance.
Because \(e_T\neq0\), however, the periapsis direction of the target body's orbit remains physically meaningful.
It determines the target-body radial distance along the two opposite directions of the line of nodes between the projectile's orbital plane and the reference plane.
It therefore continues to affect the corresponding local orbital-intersection geometry.

In this coplanar eccentric limit, the longitude of periapsis of the target body's orbit is introduced as
\begin{equation}
\varpi_T
=
\Omega_T+\omega_T,
\end{equation}
which specifies the physical orientation of its periapsis direction within the reference plane.
At the start \(\tau_{n-1}\) of the \(n\)th Kozai cycle, the angular position of the projectile's ascending-node direction relative to the target body's periapsis direction is described by
\begin{equation}
\psi_{n-1}
=
\Omega_{n-1}
-
\varpi_{T,n-1}.
\end{equation}
Combining
\begin{equation}
\Delta\Omega_{n-1}
=
\Omega_{n-1}
-
\Omega_{T,n-1}
\end{equation}
and
\begin{equation}
\varpi_{T,n-1}
=
\Omega_{T,n-1}
+
\omega_{T,n-1}
\end{equation}
gives
\begin{equation}
\psi_{n-1}
=
\Delta\Omega_{n-1}
-
\omega_{T,n-1}.
\end{equation}

Thus, for \(e_T\neq0\) and \(i_T=0\), the two cycle-start orientation variables \(\Delta\Omega_{n-1}\) and \(\omega_{T,n-1}\) no longer represent two independent geometrical degrees of freedom.
They affect the collision geometry during the subsequent complete Kozai cycle only through their combination \(\psi_{n-1}\).
Accordingly, the conditional single-cycle collision frequency reduces from
\begin{equation}
K(\mathbf{x}_{n-1})
=
K(\Delta\Omega_{n-1},\omega_{T,n-1})
\end{equation}
to
\begin{equation}
K(\psi_{n-1}).
\end{equation}

For the integration variables \((\Delta\Omega',\omega_T')\), define
\begin{equation}
\psi'
=
\Delta\Omega'
-
\omega_T'.
\end{equation}
The general two-dimensional orientation space therefore reduces in this limit to a one-dimensional orientation space described by the angular variable \(\psi'\), and the conditional single-cycle collision frequency can be written as
\begin{equation}
K(\psi').
\end{equation}
Substituting this relation into equation~\eqref{eq:long-term-orientation-average} and using the \(2\pi\)-periodicity of \(K(\psi')\) gives
\begin{equation}
\Gamma
=
\frac{1}{(2\pi)^2}
\int_0^{2\pi}
\int_0^{2\pi}
K(\Delta\Omega'-\omega_T')
\,{\rm d}\Delta\Omega'
\,{\rm d}\omega_T'
=
\frac{1}{2\pi}
\int_0^{2\pi}
K(\psi')
\,{\rm d}\psi'.
\end{equation}
Thus, in this special limit, the average over the two-dimensional orientation space reduces to an average over the single variable \(\psi'\).

In this coplanar eccentric limit, the rate of change of the target body's longitude of periapsis is defined as
\begin{equation}
\dot{\varpi}_T
\equiv
\frac{{\rm d}\varpi_T}{{\rm d}t}.
\end{equation}
When the general orientation parametrization introduced above is continued to the coplanar limit \(i_T=0\), the corresponding rates satisfy
\begin{equation}
\dot{\varpi}_T
=
\dot{\Omega}_T
+
\dot{\omega}_T.
\end{equation}
Over the physical duration \(T_{\rm K}\) of one complete Kozai cycle, the change in the target body's longitude of periapsis is
\begin{equation}
\delta_{\varpi_T}
=
\dot{\varpi}_T T_{\rm K}.
\end{equation}
The angular difference \(\psi\) at successive Kozai-cycle starts therefore satisfies
\begin{equation}
\psi_n
=
\psi_{n-1}
+
\delta_\psi
\pmod{2\pi},
\end{equation}
where
\begin{equation}
\delta_\psi
=
\delta_\Omega
-
\dot{\varpi}_T T_{\rm K}.
\end{equation}

For this reduced one-dimensional problem, replacing the long-term cycle average by a uniform average over \(\psi'\) requires only that \(\delta_\psi\) be incommensurate with \(2\pi\).
Equivalently, there are no integers \(m\neq0\) and \(q\) such that
\begin{equation}
m\,\delta_\psi
=
2\pi q.
\end{equation}
Under this condition, the successive cycle-start values \(\psi_{n-1}\) are equidistributed over the corresponding one-dimensional orientation space in the long-term limit.
The sequence of cycle-start orientations generated by the secular evolution therefore reproduces the same long-term statistics as the uniform average over \(\psi'\) derived above.

The uniform average over \(\psi'\) can be recast by changing the integration variable from \(\psi'\) to the radial distance \(r_T\) of the target body's orbit.
For a prescribed cycle-start orientation \(\psi'\), the projectile's ascending-node direction at that cycle start is taken as the reference direction.
The quantity \(r_T\) denotes the radial distance from the central body to the target body's orbit along this direction.
Because \(\psi'\) specifies the angular position of this direction relative to the target body's periapsis direction,
\begin{equation}
r_T
=
\frac{
a_T(1-e_T^2)
}{
1+e_T\cos\psi'
}.
\end{equation}

Let the periapsis and apoapsis distances of the target body's orbit be
\begin{equation}
r_1
=
a_T(1-e_T),
\qquad
r_2
=
a_T(1+e_T),
\end{equation}
respectively.

For any radial distance satisfying
\begin{equation}
r_1<r_T<r_2,
\end{equation}
there are two distinct angular positions \(\psi'\) that correspond to the same value of \(r_T\).
From
\begin{equation}
\cos\psi'
=
\frac{
a_T(1-e_T^2)/r_T-1
}{
e_T
},
\end{equation}
these two angular positions may be denoted by
\begin{equation}
\psi'_+(r_T),
\qquad
\psi'_-(r_T).
\end{equation}
They have the same value of \(\cos\psi'\) but opposite signs of \(\sin\psi'\).

Changing the integration variable from \(\psi'\) to \(r_T\) gives
\begin{equation}
\left|
\frac{{\rm d}\psi'}{{\rm d}r_T}
\right|
=
\frac{
a_T\sqrt{1-e_T^2}
}{
r_T
\sqrt{
(r_T-r_1)(r_2-r_T)
}
}.
\end{equation}
Because every \(r_T\) in the interval \(r_1<r_T<r_2\) corresponds to two angular positions and \(\psi'\) is uniformly distributed over \([0,2\pi)\), the normalized radial weight induced by this change of variables is
\begin{equation}
W_r(r_T)
=
\frac{
a_T\sqrt{1-e_T^2}
}{
\pi r_T
\sqrt{
(r_T-r_1)(r_2-r_T)
}
},
\end{equation}
with
\begin{equation}
\int_{r_1}^{r_2}
W_r(r_T)
\,{\rm d}r_T
=
1.
\end{equation}

Here, \(W_r(r_T)\) is induced solely by the change of variables from the uniformly distributed orientation variable \(\psi'\) to \(r_T\).
It does not represent the fraction of time that the target body spends at different radial distances along its Keplerian orbit.

The two angular positions \(\psi'_+(r_T)\) and \(\psi'_-(r_T)\) corresponding to the same radial distance generally have different local directions of motion.
Their conditional single-cycle collision frequencies therefore need not be identical.
Accordingly,
\begin{equation}
K_+(r_T)
=
K\!\left[\psi'_+(r_T)\right],
\qquad
K_-(r_T)
=
K\!\left[\psi'_-(r_T)\right].
\end{equation}
For a given \(r_T\), the mean conditional single-cycle collision frequency over these two angular positions is then defined as
\begin{equation}
K_r(r_T)
=
\frac{1}{2}
\left[
K_+(r_T)
+
K_-(r_T)
\right].
\end{equation}

The uniform average over \(\psi'\) can therefore be written equivalently as the weighted radial average
\begin{equation}
\Gamma
=
\int_{r_1}^{r_2}
W_r(r_T)
K_r(r_T)
\,{\rm d}r_T.
\end{equation}
Hence,
\begin{equation}
\frac{1}{2\pi}
\int_0^{2\pi}
K(\psi')
\,{\rm d}\psi'
=
\int_{r_1}^{r_2}
W_r(r_T)
K_r(r_T)
\,{\rm d}r_T.
\end{equation}

Thus, for \(e_T\neq0\) and \(i_T=0\), the average over \(\psi'\) can be represented equivalently as a weighted average over the radial distance \(r_T\).

In the general dynamical setup adopted here, the periapsis direction of the target body's orbit is allowed to evolve at a prescribed rate.
Under the additional condition
\begin{equation}
\dot{\varpi}_T=0,
\end{equation}
the periapsis direction of the target body's eccentric orbit remains fixed in the reference plane.
Under this condition, the weighted average over \(r_T\) corresponds to the integral form adopted by \citet{pokorny2013} for an eccentric target-body orbit in the reference plane.
The uniform average over \(\psi'\) and the weighted radial average over \(r_T\) are therefore two equivalent representations of the same long-term mean collision frequency using different integration variables.

\section{Numerical results and validation}
\label{sec:results}

To validate the long-term mean collision-frequency framework developed above, four representative target-body orbital configurations, denoted Cases 1--4, are considered.
Cases 1, 2, and 3 correspond to \(e_T=0,\ i_T=0\), \(e_T=0,\ i_T\neq0\), and \(e_T\neq0,\ i_T=0\), respectively, and are used to test the theoretical reductions derived in Section~\ref{sec:limits}.
Case 4 considers the general eccentric and inclined target-body orbit with \(e_T\neq0\) and \(i_T\neq0\), for which the theoretical result is compared with independent direct dynamical integrations performed with REBOUND.

All four cases use the same Sun--Jupiter secular dynamical model for the collision-frequency calculations, with the Sun as the central body, Jupiter as the perturbing body, and Jupiter's orbital plane as the reference plane.
Jupiter is placed on a circular orbit with semimajor axis \(a_P=5.2\,{\rm AU}\), zero eccentricity, and zero inclination.
\begin{table}[!b]
\centering
\small
\caption{Main parameters adopted in the four validation cases.}
\label{tab:validation-parameters}
\renewcommand{\arraystretch}{1.12}
\begin{tabular*}{\textwidth}{@{\extracolsep{\fill}}ccccc}
\toprule
Parameter
& Case 1
& Case 2
& Case 3
& Case 4
\\
\midrule
\multicolumn{5}{c}{\textit{Projectile parameters}}
\\
\addlinespace[0.3ex]
\(a\,[{\rm AU}]\)
& 1.4
& 1.4
& 0.9
& 0.9
\\
\(e_0\)
& 0.2
& 0.2
& 0.1
& 0.1
\\
\(i_0\)
& \(65^\circ\)
& \(65^\circ\)
& \(55^\circ\)
& \(55^\circ\)
\\
\(\omega_0\)
& \(0^\circ\)
& \(20^\circ\)
& \(0^\circ\)
& \(0^\circ\)
\\
\(\Omega_0\)
& \(0^\circ\)
& \(0^\circ\)
& \(0^\circ\)
& \(0^\circ\)
\\
\addlinespace[0.6ex]
\multicolumn{5}{c}{\textit{Target-body parameters}}
\\
\addlinespace[0.3ex]
\(a_T\,[{\rm AU}]\)
& 1.0
& 1.0
& 0.3871
& 0.3871
\\
\(e_T\)
& 0
& 0
& 0.2056
& 0.2056
\\
\(i_T\)
& \(0^\circ\)
& \(10^\circ\)
& \(0^\circ\)
& \(10^\circ\)
\\
\(\Omega_{T,0}\)
& --
& \(0^\circ\)
& --
& \(0^\circ\)
\\
\(\omega_{T,0}\)
& --
& --
& --
& \(0^\circ\)
\\
\(\varpi_{T,0}\)
& --
& --
& \(0^\circ\)
& --
\\
\(R_T\,[{\rm AU}]\)
& \(4.26\times10^{-4}\)
& \(8.527\times10^{-4}\)
& \(1.63\times10^{-4}\)
& \(2.445\times10^{-4}\)
\\
\(\dot{\Omega}_T\,[{\rm yr}^{-1}]\)
& --
& \(0\)
& --
& \(-2\pi/(1.5\times10^{5})\)
\\
\(\dot{\omega}_T\,[{\rm yr}^{-1}]\)
& --
& --
& --
& \(-2\pi/(3.0\times10^{5})\)
\\
\(\dot{\varpi}_T\,[{\rm yr}^{-1}]\)
& --
& --
& \(0\)
& --
\\
\addlinespace[0.6ex]
\multicolumn{5}{c}{\textit{Validation comparison}}
\\
\addlinespace[0.3ex]
Comparison
& \citet{vokrouhlicky2012}
& \citet{Liang2026}
& \citet{pokorny2013}
& REBOUND integrations
\\
\bottomrule
\end{tabular*}

\vspace{0.6ex}
\begin{minipage}{0.97\textwidth}
\footnotesize
\textit{Note.}
A dash denotes a quantity that is not used in the corresponding case; for a circular or zero-inclination orbit, the relevant periapsis or nodal orientation, and hence its associated orientation rate, has no independent physical significance.
\end{minipage}
\end{table}
The main parameters adopted for the four cases are listed in Table~\ref{tab:validation-parameters}.

\subsection{Validation for special target-body orbits}
\label{sec:results-limits}

The long-term mean collision frequencies obtained with the present framework are compared with the corresponding reference results in Table~\ref{tab:limit-comparison}.
The relative difference between the present result and the corresponding reference value is defined as
\begin{equation}
\epsilon
=
\frac{
\left|
\Gamma-\Gamma_{\rm ref}
\right|
}{
\Gamma_{\rm ref}
}
\times100\%.
\label{eq:validation-relative-difference}
\end{equation}

\begin{table}[!b]
\centering
\small
\caption{Long-term mean collision frequencies for the three special target-body orbital configurations and comparison with the corresponding reference results.}
\label{tab:limit-comparison}
\begin{tabular}{cccc}
\toprule
Case
&
Present result
&
Reference result
&
Relative difference
\\
&
\(\Gamma\,[{\rm yr}^{-1}]\)
&
\(\Gamma_{\rm ref}\,[{\rm yr}^{-1}]\)
&
\(\epsilon\)
\\
\midrule
Case 1
&
\(1.0661\times10^{-7}\)
&
\(1.0634\times10^{-7}\)
&
\(0.2508\%\)
\\
Case 2
&
\(3.8069\times10^{-7}\)
&
\(3.8000\times10^{-7}\)
&
\(0.1810\%\)
\\
Case 3
&
\(7.0075\times10^{-8}\)
&
\(6.9537\times10^{-8}\)
&
\(0.7741\%\)
\\
\bottomrule
\end{tabular}

\vspace{0.5ex}
\begin{minipage}{0.96\columnwidth}
\footnotesize
\textit{Note.}
For Cases 1 and 3, the reference collision frequencies \(\Gamma_{\rm ref}\) are converted from the intrinsic collision probabilities \(p_{\rm ref}\) reported in the corresponding studies using \(\Gamma_{\rm ref}=p_{\rm ref}R_T^2\).
\end{minipage}
\end{table}

The relative differences for Cases 1, 2, and 3 are \(0.2508\%\), \(0.1810\%\), and \(0.7741\%\), respectively, and are all below \(1\%\).
Thus, the present framework not only reduces analytically to the corresponding special formulations considered in previous studies, but also yields numerical results in good agreement with the corresponding reference values.
The small remaining differences may arise from the finite numerical precision of the reference results and from differences in numerical integration and discretization.

\subsection{Direct dynamical validation for the general target-body orbit}
\label{sec:results-general}

Case~4 considers the general configuration in which the target body's orbit has both non-zero eccentricity and non-zero inclination.
Averaging over the two-dimensional orientation space as described in Section~\ref{sec:theoretical-framework} gives the long-term mean collision frequency \( \Gamma = 1.1851\times10^{-7}\,{\rm yr}^{-1}\).

To validate this result independently, direct dynamical integrations are performed with REBOUND \citep{ReinLiu2012}.
The Sun and Jupiter are included as active gravitating bodies.
The target body moves on a prescribed Keplerian elliptical orbit, while its longitude of ascending node \(\Omega_T\) and argument of periapsis \(\omega_T\) precess at the prescribed rates.
The target body's initial mean anomaly is set to \(M_{T,0}=0\).
The projectiles are treated as massless test particles and evolve under the gravity of the Sun and Jupiter.
Mutual gravitational interactions between the target body and the projectiles are neglected.

\begin{figure}
\centering
\includegraphics[width=0.75\columnwidth]{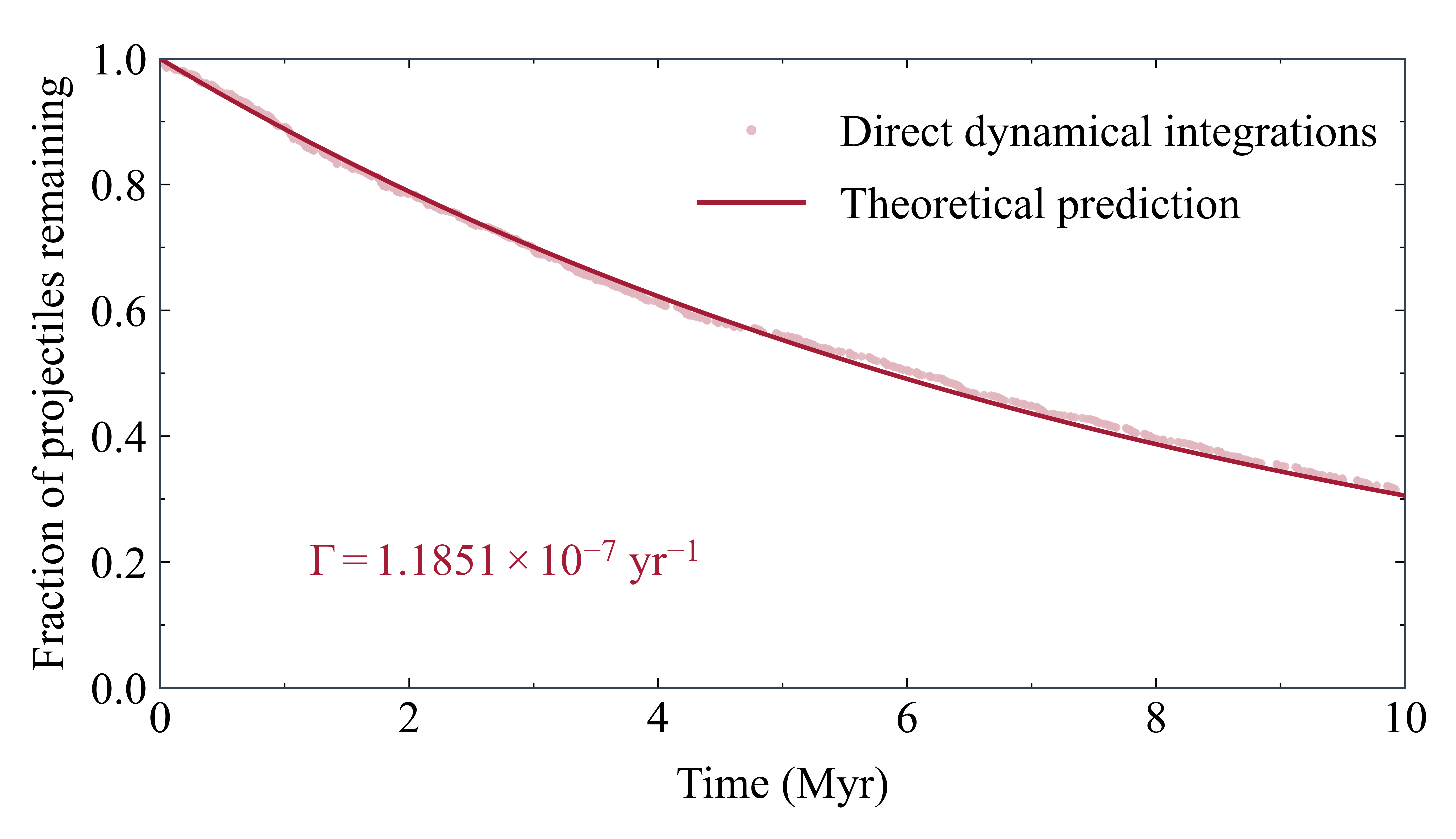}
\caption{
Fraction of projectiles remaining in Case~4 from the direct dynamical integrations and the theoretical prediction.
}
\label{fig:case4-survival}
\end{figure}

The direct integrations contain \(N_0=1000\) projectiles.
All projectiles belong to the same closed Hamiltonian level curve used for Case~4.
The projectile elements listed for Case~4 in Table~\ref{tab:validation-parameters} specify the reference secular state used to define this Hamiltonian level curve.
To avoid starting all projectiles from the same secular state on this level curve, their initial \(e\), \(i\), and \(\omega\) are sampled at different points along one complete Lidov--Kozai cycle according to its physical-time parametrization.
The projectiles therefore have the same Hamiltonian value \(H_0\) and Kozai constant \(c\), but start from different secular states along the same Hamiltonian level curve.
Their initial longitudes of ascending node \(\Omega\) and mean anomalies \(M\) are sampled independently and uniformly over \([0,2\pi)\).

Under the incommensurability conditions adopted above, selecting different initial secular states along the same closed Hamiltonian level curve does not change the long-term mean collision frequency considered here.
A proof of this property is given in Section~\ref{sec:discussion}.

The integrations use the WHFast integrator \citep{ReinTamayo2015} with a time step of \(2\,{\rm d}\)
and a total integration time of \(10^7\,{\rm yr}\).
A collision with the target body is recorded when the projectile--target centre-to-centre separation becomes smaller than \(R_T\).
Collisions are detected using REBOUND's \texttt{linetree} collision-search algorithm, and the corresponding projectile is removed once a collision is identified.

Treating the long-term mean collision frequency \(\Gamma\) as a constant removal rate in the long-term statistical sense, the predicted fraction of projectiles remaining is
\begin{equation}
S(t)
=
\exp(-\Gamma t).
\label{eq:theoretical-remaining-fraction}
\end{equation}
Let \(N_{\rm rem}(t)\) denote the number of projectiles remaining in the REBOUND integrations at time \(t\).
The fraction obtained from the direct integrations is therefore \(N_{\rm rem}(t)/N_0\).

Figure~\ref{fig:case4-survival} compares the fraction of projectiles remaining in the direct dynamical integrations performed with REBOUND with the theoretical prediction given by equation~\eqref{eq:theoretical-remaining-fraction}.
The two remain in close agreement over the full integration interval.

\section{Effect of the projectile's initial secular state along a Hamiltonian level curve}
\label{sec:discussion}

In the REBOUND direct dynamical integrations described above, the projectiles are initialized with eccentricities, inclinations, and arguments of periapsis corresponding to different positions along the same closed Hamiltonian level curve.
It is shown below that this choice of projectile initial conditions does not affect the long-term mean collision frequency.

First, consider an initial configuration at the dimensionless secular time \(\tau_0\), for which the initial eccentricity, inclination, argument of periapsis, and longitude of ascending node of the projectile are \(e_0,i_0,\omega_0\) and \(\Omega_0^{(1)}\).
The initial longitude of ascending node and argument of periapsis of the target body are
\(\Omega_{T,0}\) and \(\omega_{T,0}\).
The corresponding long-term mean collision frequency is denoted by \(\Gamma_1\).

Consider another fixed dimensionless secular time \(\tau_1\) along this evolution.
At \(\tau_1\), the eccentricity, inclination, argument of periapsis, and longitude of ascending node of the projectile are \(e(\tau_1), i(\tau_1), \omega(\tau_1)\) and \(\Omega(\tau_1)\), while the longitude of ascending node and argument of periapsis of the target body are \(\Omega_T(\tau_1)\) and \(\omega_T(\tau_1)\).
Taking \(\tau_1\) as a new starting point defines a second initial configuration, with the quantities listed above taken as its initial values.
The corresponding long-term mean collision frequency is denoted by \(\Gamma_2\).

The second initial configuration is obtained simply by shifting the starting point of the first evolution from \(\tau_0\) to \(\tau_1\).
The subsequent relative orbital evolution of the projectile and the target body is unchanged.
Such a shift of the time origin therefore does not change the long-term mean collision frequency, and hence
\begin{equation}
\Gamma_1=\Gamma_2.
\label{eq:initial-state-gamma12}
\end{equation}

Next, consider a third initial configuration.
The initial dimensionless secular time is again \(\tau_0\), while the initial eccentricity, inclination, and argument of periapsis of the projectile are taken to be
\(e(\tau_1), i(\tau_1)\) and \(\omega(\tau_1)\).
Its initial longitude of ascending node is assigned an arbitrary value \(\Omega_0^{(2)}\).
The initial longitude of ascending node and argument of periapsis of the target body remain
\(\Omega_{T,0}\) and \(\omega_{T,0}\).
The corresponding long-term mean collision frequency is denoted by \(\Gamma_3\).

The second and third initial configurations have the same initial eccentricity, inclination, and argument of periapsis of the projectile.
They therefore correspond to the same secular state on the same closed Hamiltonian level curve and have the same Hamiltonian value \(H_0\) and Kozai constant \(c\).
The two configurations differ only in their initial orientation variables, which are respectively
\begin{equation}
\left[
\Omega(\tau_1)-\Omega_T(\tau_1),
\,
\omega_T(\tau_1)
\right]
\end{equation}
and
\begin{equation}
\left[
\Omega_0^{(2)}-\Omega_{T,0},
\,
\omega_{T,0}
\right].
\end{equation}

Under the incommensurability condition considered here, the cycle-start orientations are equidistributed over the two-dimensional orientation space \(\mathcal{X}\) in the long-term limit, and this distribution is independent of the initial values of the orientation variables.
The second and third initial configurations therefore have the same long-term mean collision frequency,
\begin{equation}
\Gamma_2=\Gamma_3.
\label{eq:initial-state-gamma23}
\end{equation}

Combining equations~\eqref{eq:initial-state-gamma12} and~\eqref{eq:initial-state-gamma23} gives
\begin{equation}
\Gamma_1=\Gamma_2=\Gamma_3.
\label{eq:initial-state-invariance}
\end{equation}

Because the projectile traverses the entire closed Hamiltonian level curve during one complete Kozai cycle, \(\tau_1\) may correspond to any secular state on that level curve, while \(\Omega_0^{(2)}\) may be chosen arbitrarily.
Therefore, with the dynamical parameters otherwise fixed and under the incommensurability condition considered here, choosing different initial secular states of the projectile along the same closed Hamiltonian level curve does not change the long-term mean collision frequency.
Changing the initial longitude of ascending node of the projectile likewise does not affect this long-term mean quantity.

To further test this theoretical result, an additional numerical test is performed using the general eccentric and inclined target-body orbit of Case~4.
Ten positions are selected along the same closed Hamiltonian level curve of Case~4, equally spaced in the dimensionless secular time \(\tau\) over one complete Kozai cycle.
The eccentricity, inclination, and argument of periapsis at these positions are adopted as the corresponding initial orbital elements of the projectile, while its initial longitude of ascending node \(\Omega\) is randomly selected from \([0,2\pi)\).
All other physical parameters are kept the same as in Case~4.

\begin{figure}
\centering
\includegraphics[width=0.75\columnwidth]{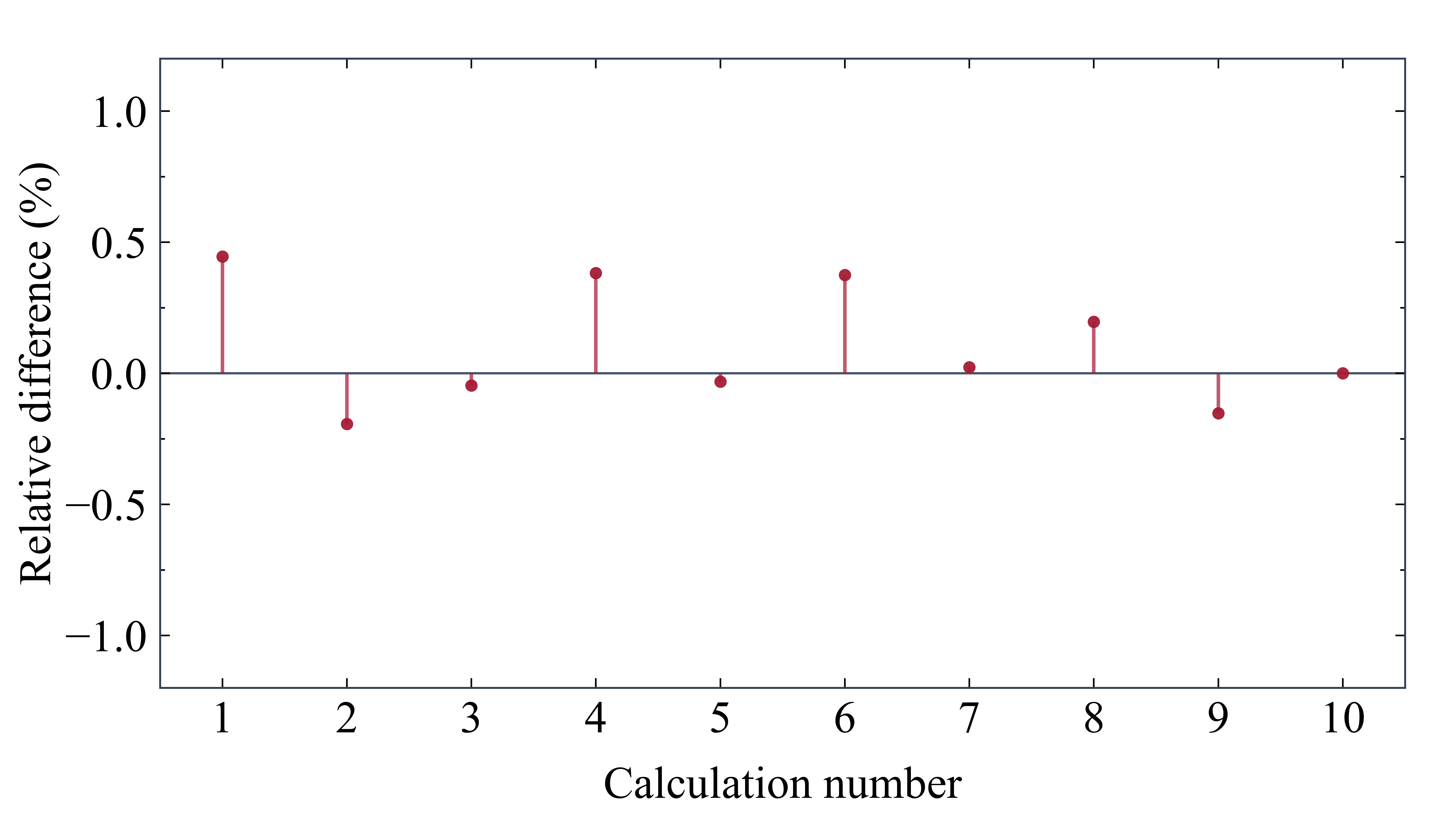}
\caption{
Relative differences in the long-term mean collision frequency for ten initial secular states along the same closed Hamiltonian level curve in Case~4, measured with respect to the original Case~4 result.
}
\label{fig:case4_hamiltonian_phase_test}
\end{figure}

Figure~\ref{fig:case4_hamiltonian_phase_test} shows the relative differences between the long-term mean collision frequencies obtained from the ten calculations and the original Case~4 result.
All absolute differences remain below \(0.5\%\), with no systematic trend among the calculations.
These results are consistent with the invariance derived above.

\section{Conclusions}

This study establishes a unified framework for calculating the long-term mean collision frequency between high-inclination projectiles undergoing Lidov--Kozai secular evolution and a target body.
The framework extends previous treatments of special target-body orbital configurations to the case in which the target body's orbit has both non-zero eccentricity and non-zero inclination, with prescribed evolution of its orbital orientation.

When the target body's orbit is circular or its orbital inclination is zero, the present framework reduces to the corresponding frameworks developed in previous studies.

In addition, it is shown that, for a fixed dynamical setup and under the incommensurability condition, the long-term mean collision frequency for a given closed Hamiltonian level curve is unchanged when the projectile is initialized at different secular states along that level curve or with different initial longitudes of ascending node.
This property provides a basis for grouping such projectiles and assigning them the same long-term mean collision frequency.
Such grouping reduces redundant calculations when collision frequencies are evaluated for projectile populations with different initial conditions.

\section*{Acknowledgements}
This work was supported by the National Natural Science Foundation of China (Nos.~12472048, 12002397 and 62388101), and the Shenzhen Science and Technology Program (Grant No.~ZDSYS20210623091808026).

\section*{Data Availability}
The data underlying this work will be made available by the corresponding author upon reasonable request.

\section*{Conflict of Interest}
The authors declare no conflicts of interest.

\bibliographystyle{plainnat}
\bibliography{paper4_references}

\appendix
\section{Calculation of the conditional single-cycle collision frequency}
\label{app:general_eccentric_inclined_target}

The calculation of the conditional single-cycle collision frequency follows the basic construction developed by \citet{Liang2026} for a target body on an inclined circular orbit.
The present treatment extends this calculation to the case in which the target body's orbit has both non-zero eccentricity and non-zero inclination.
The overall slow-variable--fast-phase decomposition remains unchanged, whereas the specific expressions for the orbital-intersection geometry and the associated physical quantities are generalized.

Section~\ref{sec:2-3} defines the conditional single-cycle collision frequency for a specified cycle-start orientation.
Denoting an arbitrary cycle-start orientation by
\begin{equation}
\mathbf{x}'
=
\left(
\Delta\Omega',
\omega_T'
\right),
\end{equation}
the corresponding frequency is
\begin{equation}
K(\mathbf{x}')
=
\frac{1}{T_{\rm orb}}
\sum_{s=\pm1}
\sum_w
\Pi_{s,w}
=
\frac{1}{T_{\rm orb}}
\sum_{s=\pm1}
\sum_w
P_{1,s,w}P_{2,s,w}.
\label{eq:app_single_cycle_frequency}
\end{equation}
For a specified \(\mathbf{x}'\), the exact orbital-intersection roots \(\tau_{s,w}\) are determined during the subsequent complete Lidov--Kozai cycle, together with the corresponding slow-variable probability factors \(P_{1,s,w}\) and fast-phase probability factors \(P_{2,s,w}\).
The root-level collision contribution is
\begin{equation}
\Pi_{s,w}
=
P_{1,s,w}P_{2,s,w}.
\label{eq:app_root_contribution}
\end{equation}
The expressions required to evaluate \(\tau_{s,w}\), \(P_{1,s,w}\), and \(P_{2,s,w}\) when the target body's orbit has both non-zero eccentricity and non-zero inclination are given below.

For a specified cycle-start orientation \(\mathbf{x}'\), the orbital evolution over the subsequent complete Lidov--Kozai cycle is determined by the dynamical model described in Section~\ref{sec:theoretical-framework}.
At an arbitrary dimensionless secular time \(\tau\), the intersection of the projectile's orbital plane with the target body's orbital plane defines the mutual line of nodes.
Its two opposite directions form the two mutual-node branches, labelled by \(s=\pm1\).
The direction of the cross product of the target body's orbital angular-momentum vector with that of the projectile is assigned to \(s=+1\), and the opposite direction to \(s=-1\).

The relative nodal longitude is
\begin{equation}
\Delta\Omega(\tau)
=
\Omega(\tau)-\Omega_T(\tau).
\label{eq:app_relative_node}
\end{equation}
The arguments of latitude associated with mutual-node branch \(s\) in the projectile and target orbital planes are denoted by \(u_s(\tau)\) and \(u_{T,s}(\tau)\), respectively:
\begin{equation}
\begin{aligned}
u_s
&=
\operatorname{atan2}\!\Bigl[
s\sin i_T\sin\Delta\Omega,\,
s\bigl(
\sin i\cos i_T
-
\cos i\sin i_T\cos\Delta\Omega
\bigr)
\Bigr],
\\
u_{T,s}
&=
\operatorname{atan2}\!\Bigl[
s\sin i\sin\Delta\Omega,\,
s\bigl(
\sin i\cos i_T\cos\Delta\Omega
-
\cos i\sin i_T
\bigr)
\Bigr].
\end{aligned}
\label{eq:app_mutual_node_arguments}
\end{equation}
Hereafter, the explicit dependence of these instantaneous quantities on \(\tau\) is omitted where no confusion can arise.

The target body's true anomaly along mutual-node branch \(s\) is
\begin{equation}
f_{T,s}
=
u_{T,s}-\omega_T.
\label{eq:app_target_true_anomaly}
\end{equation}

Using the nonsingular eccentricity variables \(k\) and \(h\) defined in the main text, and setting \(g=\sqrt{1-e^2}\), the radial distances from the central body to the projectile's orbit and the target body's orbit along mutual-node branch \(s\) are
\begin{equation}
r_s
=
\frac{ag^2}
{1+k\cos u_s+h\sin u_s},
\label{eq:app_projectile_radius}
\end{equation}
and
\begin{equation}
r_{T,s}
=
\frac{
a_T(1-e_T^2)
}{
1+e_T\cos f_{T,s}
}.
\label{eq:app_target_radius}
\end{equation}
The corresponding signed radial-distance difference is defined as
\begin{equation}
G_s
=
r_s-r_{T,s}
=
\frac{ag^2}
{1+k\cos u_s+h\sin u_s}
-
\frac{
a_T(1-e_T^2)
}{
1+e_T\cos(u_{T,s}-\omega_T)
}.
\label{eq:app_Gs_eccentric_target}
\end{equation}
For \(e_T=0\), \(r_{T,s}=a_T\), and equation~\eqref{eq:app_Gs_eccentric_target} reduces to the signed radial-distance difference used by \citet{Liang2026} for a target body on an inclined circular orbit.

The exact orbital-intersection roots \(\tau_{s,w}\) satisfy
\begin{equation}
G_s(\tau_{s,w})=0,
\label{eq:app_exact_root_eccentric_target}
\end{equation}
where \(w\) labels the distinct roots on mutual-node branch \(s\) during one complete Lidov--Kozai cycle.

The quantities \(\alpha_s\) and \(\alpha_{T,s}\) denote the directed angles from mutual-node branch \(s\) to the local tangents to the instantaneous Keplerian orbits of the projectile and target body, respectively.
They satisfy
\begin{equation}
\cot\alpha_s
=
\frac{
k\sin u_s-h\cos u_s
}{
1+k\cos u_s+h\sin u_s
},
\label{eq:app_cot_alpha_projectile}
\end{equation}
and
\begin{equation}
\cot\alpha_{T,s}
=
\frac{
e_T\sin f_{T,s}
}{
1+e_T\cos f_{T,s}
}.
\label{eq:app_cot_alpha_target}
\end{equation}
For \(e_T=0\), \(\cot\alpha_{T,s}=0\).

The signed mutual inclination associated with mutual-node branch \(s\) is denoted by \(I_s\), whose cosine is
\begin{equation}
\cos I_s
=
\cos i\cos i_T
+
\sin i\sin i_T\cos\Delta\Omega.
\label{eq:app_signed_mutual_inclination}
\end{equation}
The sign convention follows \citet{Liang2026}: \(I_s>0\) when the projectile crosses the target body's orbital plane in the direction of the target body's orbital angular-momentum vector, and \(I_s<0\) when it crosses in the opposite direction.

Following the local straight-line treatment near a mutual node considered by \citet{Greenberg1982}, the minimum separation \(D_s\) between the two local tangent lines satisfies
\begin{equation}
D_s
\simeq
B_s|G_s|,
\label{eq:app_Ds_general}
\end{equation}
where the dimensionless geometrical projection factor \(B_s\) is
\begin{equation}
B_s
=
\frac{
|\sin I_s|
}{
\left[
\cot^2\alpha_s
+
\cot^2\alpha_{T,s}
+
\sin^2 I_s
-
2\cot\alpha_s\cot\alpha_{T,s}\cos I_s
\right]^{1/2}
}.
\label{eq:app_Bs_general}
\end{equation}
For \(e_T=0\), \(\cot\alpha_{T,s}=0\), and equation~\eqref{eq:app_Bs_general} reduces to the geometrical projection factor used by \citet{Liang2026} for a target body on an inclined circular orbit.

Near an exact orbital-intersection root \(\tau_{s,w}\),
\begin{equation}
G_s(\tau)
\simeq
\left.
\frac{{\rm d}G_s}{{\rm d}\tau}
\right|_{\tau=\tau_{s,w}}
\left(
\tau-\tau_{s,w}
\right).
\label{eq:app_Gs_linearization}
\end{equation}
When the target body's orbit has both non-zero eccentricity and non-zero inclination,
\begin{equation}
G_s
=
G_s
\left(
k,h,\Delta\Omega,\omega_T
\right).
\end{equation}
By the chain rule, the total derivative of \(G_s\) along the time-ordered curve is
\begin{equation}
\frac{{\rm d}G_s}{{\rm d}\tau}
=
\frac{\partial G_s}{\partial k}
\frac{{\rm d}k}{{\rm d}\tau}
+
\frac{\partial G_s}{\partial h}
\frac{{\rm d}h}{{\rm d}\tau}
+
\frac{\partial G_s}{\partial\Delta\Omega}
\frac{{\rm d}\Delta\Omega}{{\rm d}\tau}
+
\frac{\partial G_s}{\partial\omega_T}
\frac{{\rm d}\omega_T}{{\rm d}\tau}.
\label{eq:app_dGs_dtau_general}
\end{equation}
Here
\begin{equation}
\frac{{\rm d}\Delta\Omega}{{\rm d}\tau}
=
\frac{{\rm d}\Omega}{{\rm d}\tau}
-
\dot{\Omega}_T
\frac{{\rm d}t}{{\rm d}\tau},
\qquad
\frac{{\rm d}\omega_T}{{\rm d}\tau}
=
\dot{\omega}_T
\frac{{\rm d}t}{{\rm d}\tau}.
\label{eq:app_orientation_derivatives}
\end{equation}

At \(\tau_{s,w}\), let
\begin{equation}
B_{s,w}
=
B_s(\tau_{s,w}).
\end{equation}
The linear approximation gives the candidate half-width of the local time window as
\begin{equation}
\Delta\tau_{s,w}^{\rm lin}
=
\frac{
R
}{
B_{s,w}
\left|
\left.
{\rm d}G_s/{\rm d}\tau
\right|_{\tau=\tau_{s,w}}
\right|
}.
\label{eq:app_window_linear_half_width}
\end{equation}
The final window boundaries \(\tau_{s,w}^{-}\) and \(\tau_{s,w}^{+}\) are determined using the window-construction procedure of \citet{Liang2026}.
The boundaries estimated from equation~\eqref{eq:app_window_linear_half_width} are checked against the local separation \(D_s(\tau)\simeq B_s(\tau)|G_s(\tau)|\) and, when necessary, refined through an adaptive boundary search.

The physical duration of the local time window is
\begin{equation}
\Delta t_{s,w}
=
\int_{\tau_{s,w}^{-}}^{\tau_{s,w}^{+}}
\frac{{\rm d}t}{{\rm d}\tau}
\,{\rm d}\tau
=
\int_{\tau_{s,w}^{-}}^{\tau_{s,w}^{+}}
\frac{16}
{\gamma_\star g(\tau)}
\,{\rm d}\tau,
\label{eq:app_window_physical_duration}
\end{equation}
and the corresponding slow-variable probability factor is
\begin{equation}
P_{1,s,w}
=
\frac{\Delta t_{s,w}}{T_{\rm K}},
\label{eq:app_P1_definition}
\end{equation}
where \(T_{\rm K}\) is the physical duration of one complete Lidov--Kozai cycle.

At the exact orbital-intersection configuration corresponding to \(\tau_{s,w}\), the instantaneous velocities of the projectile and target body are denoted by \(\boldsymbol v_{p,s,w}\) and \(\boldsymbol v_{T,s,w}\), respectively.
The relative velocity is
\begin{equation}
\boldsymbol U_{s,w}
=
\boldsymbol v_{p,s,w}
-
\boldsymbol v_{T,s,w},
\qquad
U_{s,w}
=
|\boldsymbol U_{s,w}|.
\label{eq:app_relative_velocity}
\end{equation}
The target body's speed is
\begin{equation}
V_{T,s,w}
=
|\boldsymbol v_{T,s,w}|,
\end{equation}
and \(U_{\perp,s,w}\) denotes the magnitude of the component of \(\boldsymbol U_{s,w}\) perpendicular to the local tangent direction of the target body's orbit.

The reference circular speed associated with the target body's semimajor axis \(a_T\) is
\begin{equation}
V_{0,T}
=
\sqrt{\frac{\mu_0}{a_T}}.
\label{eq:app_V0T}
\end{equation}
Following the collision-probability result of \citet{Wetherill1967}, as recast for an eccentric target orbit by \citet{pokorny2013}, the fast-phase probability factor can be written as
\begin{equation}
P_{2,s,w}
=
\frac{R}{4a_T}
\frac{V_{0,T}}{V_{T,s,w}}
\frac{U_{s,w}}{U_{\perp,s,w}}.
\label{eq:app_P2_velocity_form}
\end{equation}
The factor \(V_{0,T}/V_{T,s,w}\) accounts for the variation of the target body's local orbital speed with position along its eccentric Keplerian orbit.

The common radial distance at this exact orbital-intersection configuration is
\begin{equation}
r_{s,w}
=
r_s(\tau_{s,w})
=
r_{T,s}(\tau_{s,w}).
\label{eq:app_crossing_radius}
\end{equation}
The dimensionless squared speeds of the projectile and target body, normalized by \(V_{0,T}^2\), are defined as
\begin{equation}
\mathcal V_{p,s,w}
=
a_T
\left(
\frac{2}{r_{s,w}}
-
\frac{1}{a}
\right),
\qquad
\mathcal V_{T,s,w}
=
a_T
\left(
\frac{2}{r_{s,w}}
-
\frac{1}{a_T}
\right).
\label{eq:app_dimensionless_speeds}
\end{equation}

The angle between the two local velocity directions is denoted by \(\chi_{s,w}\) and satisfies
\begin{equation}
\cos\chi_{s,w}
=
\frac{
\cot\alpha_{s,w}\cot\alpha_{T,s,w}
+
\cos I_{s,w}
}{
\left[
\left(
1+\cot^2\alpha_{s,w}
\right)
\left(
1+\cot^2\alpha_{T,s,w}
\right)
\right]^{1/2}
},
\label{eq:app_cos_chi_scalar}
\end{equation}
where
\begin{equation}
\alpha_{s,w}
=
\alpha_s(\tau_{s,w}),
\qquad
\alpha_{T,s,w}
=
\alpha_{T,s}(\tau_{s,w}),
\qquad
I_{s,w}
=
I_s(\tau_{s,w}).
\end{equation}

The relative speed and its component perpendicular to the local tangent direction of the target body's orbit satisfy
\begin{equation}
\begin{aligned}
\frac{U_{s,w}^2}{V_{0,T}^2}
&=
\mathcal V_{p,s,w}
+
\mathcal V_{T,s,w}
-
2
\left(
\mathcal V_{p,s,w}
\mathcal V_{T,s,w}
\right)^{1/2}
\cos\chi_{s,w},
\\
\frac{U_{\perp,s,w}^2}{V_{0,T}^2}
&=
\mathcal V_{p,s,w}
\left(
1-\cos^2\chi_{s,w}
\right).
\end{aligned}
\label{eq:app_relative_velocity_scalars}
\end{equation}
The equivalent scalar form of the fast-phase probability factor is
\begin{equation}
P_{2,s,w}
=
\frac{R}{4a_T}
\left[
\frac{
\mathcal V_{p,s,w}
+
\mathcal V_{T,s,w}
-
2
\left(
\mathcal V_{p,s,w}
\mathcal V_{T,s,w}
\right)^{1/2}
\cos\chi_{s,w}
}{
\mathcal V_{p,s,w}
\mathcal V_{T,s,w}
\left(
1-\cos^2\chi_{s,w}
\right)
}
\right]^{1/2}.
\label{eq:app_P2_scalar_general}
\end{equation}
For \(e_T=0\), equation~\eqref{eq:app_P2_velocity_form}, or equivalently equation~\eqref{eq:app_P2_scalar_general}, reduces directly to the compact form of the fast-phase probability factor used by \citet{Liang2026} for a target body on an inclined circular orbit.

Substitution of \(\tau_{s,w}\), \(P_{1,s,w}\), and \(P_{2,s,w}\) into equation~\eqref{eq:app_single_cycle_frequency} gives the conditional single-cycle collision frequency \(K(\mathbf{x}')\) for the specified cycle-start orientation.
\label{lastpage}
\end{document}